\documentclass[nobibnotes,nofootinbib,superscriptaddress,twocolumn,prb,showpacs]{revtex4}
\usepackage{amsmath}
\usepackage{amssymb}
\usepackage{epsfig}
\usepackage{xcolor}
\usepackage{ulem}
\usepackage{bm}

\begin{document}

\title{Electron-phonon Interaction in Non-polar Quantum Dots Induced by the Amorphous Polar Environment }

\author{A.N.~Poddubny}\email{poddubny@coherent.ioffe.ru}
\affiliation{A.F.~Ioffe Physico-Technical Institute, 194021
St.~Petersburg, Russia}

\author{S. V. Goupalov}
\affiliation{A.F.~Ioffe Physico-Technical Institute, 194021
St.~Petersburg, Russia}
\affiliation{Department of Physics, Jackson State University, Jackson, Mississippi 39217, USA
}

\author{V.I.~Kozub}
\affiliation{A.F.~Ioffe Physico-Technical Institute, 194021
St.~Petersburg, Russia}
\author{I. N. Yassievich}
\affiliation{A.F.~Ioffe Physico-Technical Institute, 194021
St.~Petersburg, Russia}
%73.21.La, 78.67.Hc -- quantum dots
%63.20.kd Phonon-electron interactions 
%66.30.hh Glasses 
%61.43.Fs Glasses 
%71.38.-k Polarons and electron-phonon interactions
 \begin{abstract}
 We propose a mechanism of energy relaxation for  carriers confined in a  non-polar quantum dot
 surrounded by an amorphous polar environment. The carrier transitions are
 due to their interaction with the  oscillating  electric field induced by the local vibrations
  in the surrounding amorphous medium.  We demonstrate that this mechanism controls energy relaxation
  for electrons in Si nanocrystals embedded in a SiO$_2$
matrix, where conventional mechanisms of electron-phonon interaction are not efficient.
 \end{abstract}
\maketitle

\def\dfrac{\displaystyle\frac}
\newcommand{\Sp}{\mathop{\rm Sp}\nolimits}
\newcommand{\St}{\mathop{\rm St}\nolimits}
\newcommand{\erf}{\mathop{\rm erf}\nolimits}
\newcommand{\arch}{\mathop{\rm Arch}\nolimits}
\newcommand{\floor}{\mathop{\rm floor}\nolimits}
\newcommand {\kp}{k_{\perp}}
\newcommand {\vkp}{\bm k_{\perp}}
\newcommand{\nn}{\nonumber}
\newcommand{\bra}{\langle}
\newcommand{\ket}{\rangle}
\newcommand{\eo}{\varepsilon_{\rm out}}
\newcommand{\ei}{\varepsilon_{\rm in}}
\newcommand{\grad}{\mathop\mathrm {grad}\nolimits}
\renewcommand{\div}{\mathop\mathrm {div}\nolimits}
\renewcommand{\phi}{\varphi}
%\newcommand{\nix}[1]{}
%\oddsidemargin=0cm \textwidth=17cm

% \title{Radiative and nonradiative relaxation of carriers confined in Si nanocrystals}
%-------------------------------------------------------------------------------------------
\section{Introduction}
%-------------------------------------------------------------------------------------------
Fr\"ohlich mechanism of electron-phonon coupling is of key importance
for carrier energy relaxation in bulk polar semiconductors as well as in
heterostructures formed by different polar
materials~\onlinecite{Ridley,Lassnig,Klein,Roca}. One of the examples of
such heterostructures is provided by semiconductor quantum dots
of polar material in a crystalline polar
environment~\onlinecite{Klein,Roca}.
Recently carrier dynamics in heterostructures combining both polar and
non-polar constituents, such as Si nanocrystals
embedded into a silica glass matrix, has attracted much attention~\onlinecite{Kenyon,Pavesi,Timmerman}.
Therefore, it becomes
interesting to analyze if the polar
environment can induce Fr\"ohlich-like interaction
within a non-polar inclusion. However, the electric field associated with a
wave of the LO-type propagating
in a polar crystal cannot penetrate through
the interface with a non-polar inclusion, since the dielectric permittivity, $\varepsilon(\omega)$,
vanishes at the frequency of
the LO-phonon mode $\omega=\omega_{LO}$. On the other hand, it is well
known that amorphous media possess high-frequency
vibrational eigenmodes of strongly localized
character~\onlinecite{taraskin1,taraskin2,olig}.
In this work we will show that the local modes in a polar
amorphous media can induce an alternating electric field even within a
non-polar inclusion and thus provide an
efficient channel for relaxation of electronic excitations of the inclusion.
Such a process can be responsible for electron
relaxation in Si nanocrystals  in SiO$_2$.
We expect this relaxation channel to be predominant as conduction-band
electron interaction with optical phonons via deformation potential is
forbidden in silicon by the crystal symmetry~\onlinecite{bp}.

The rest of the paper is organized as follows. In Sec.~\ref{sec:electrostatics}
we describe a simplified model of a non-polar spherical quantum dot surrounded by a polar environment and derive equations
for the time of carrier relaxation induced by local vibrations in the environment.
In Sec.~\ref{sec:discussion} we estimate the parameters characterizing local vibrations in a polar glass. In Sec.~\ref{sec:calculation}
we present a calculation using the values of parameters describing Si nanocrystals in a SiO$_2$ matrix. Sec.~\ref{sec:conclusions} is reserved for conclusions. Some auxiliary derivations are given in the Appendix.
\begin{figure}[h]
 \includegraphics[width=0.2\textwidth]{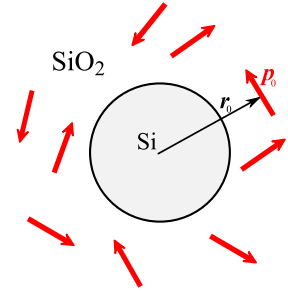}
 \caption{\label{ hedgehog}
    Schematic illustration of the Si QD surrounded by randomly oriented
    dipoles in SiO$_2$ matrix. The radius-vector $\bm r_0$ is shown for
    one of the dipoles $\bm p_0$.
 }
\end{figure}

\section{Interaction of confined electrons with local vibrations of a polar glass}\label{sec:electrostatics}

Let us consider a spherical semiconductor nanocrystal embedded into a
polar glass matrix. We will characterize the $j$th local vibration mode in the glass
by the oscillating dipole moment
\[
{\bm p}^{(j)}_0(t)=\sum\limits_i e_i^{(j)} \, {\bm u}_i^{(j)}(t) \,,
\]
where $e_i^{(j)}$ and ${\bm u}_i^{(j)}(t)$ are the charge and the displacement of
the $i$-th ion participating in the $j$-th local vibration.
The system under study is sketched in Fig.~1, where a random orientation
of the dipoles is assumed.

According to electrostatic approximation (see
Appendix) the potential inside the nanocrystal
of the radius $R$ and dielectric
constant $\ei$ induced by a point dipole source with polarization $\bm
p^{(j)}_0\delta(\bm r-\bm r_0^{(j)})$, positioned in the matrix with dielectric
constant $\eo$, can be
represented as
\begin{equation}\label{phi}
 \phi\left(\bm r,{\bm r}_0^{(j)},{\bm p}_0^{(j)}\right)=\sum\limits_{JM}\frac{\alpha_Jr^J}{\left[r_0^{(j)} \right]^{J+2}}
 \left[ \bm p_0^{(j)}\cdot\bm Y_{JM}^{J+1} \left( {\bm n}_0^{(j)} \right) \right]^*\:,
\end{equation}
where
\begin{equation}
 \alpha_J=\frac{4\pi\sqrt{(J+1)(2J+1)}}{\ei J+\eo(J+1)},\: {\bm n}_0^{(j)}=\frac{{\bm r}_0^{(j)}}{r_0^{(j)}} \,,
\end{equation}
and $\bm Y_{JM}^{L}(\bm n)$ are the vector spherical harmonics introduced in
Ref.~\onlinecite{vmk}.

Neglecting retardation, one can use the Fermi golden rule to calculate
the rate of  intraband transitions of the electron confined within
the spherical nanocrystal under the influence of the
quasistatic electric field induced by the local vibrations.
At low temperatures \cite{lowT} this rate is given by
\begin{equation}\begin{split}
\label{fgl}
\frac
 1\tau=\frac{2\pi}{\hbar}\sum\limits_j       |\bra \psi_f |e\phi(\bm r_0^{(j)}&,{\bm p}_0^{(j)})
  |\psi_i \ket|^2 \times\\ \, &\delta(E_i-E_f-\hbar\omega_{\rm ph}^{(j)}) \,,
  \end{split}
\end{equation}
where summation runs over all the local modes of the matrix,
$E_i$ and $E_f$ are, respectively, the energies of the initial
$|\psi_i \rangle$ and the final $|\psi_f \rangle$
states of the confined electron.
Neglecting electron tunneling outside the sphere, one can write the transition
matrix element as
\begin{equation}\begin{split}
 &\bra \psi_f |e\phi({\bm r}_0^{(j)},{\bm p}_0^{(j)}) |\psi_i \ket=\\ &e\sum\limits_{JM}
\frac{\alpha_J}{ \left[ r_0^{(j)} \right]^{J+2}} \bra \psi_f|r^JY_{JM}|\psi\ket
\left[ \bm p_0^{(j)}\cdot\bm Y_{JM}^{J+1} \left( {\bm n}_0^{(j)} \right)
\right]^*\:.
\end{split}
\end{equation}

Introducing the
vibration distribution function as
\[
P({\bm r}_0, {\bm p},\omega) \equiv
P_0 \, \frac{1}{4 \pi} \frac{\delta(p-p_0)}{p_0^2} \, \rho (\omega)
\]
(where $P_0 =\rm const$)
summation over vibrational modes
in Eq.~(\ref{fgl}) can be reduced to
\begin{equation}\label{tau}\begin{split}
 \frac
 1\tau=\frac{2\pi}{\hbar}&\rho\left(\tfrac{E_i-E_f}{\hbar}\right)P_0\times \\
&\int d {\bm r}_0 \int \frac{d \Omega_{{\bm p}_0}}{4 \pi}
|\bra \psi_f |e\phi(\bm r_0,{\bm p}_0)
  |\psi_i \ket|^2  \,.
  \end{split}
\end{equation}
The angular integration can be done with the help of
\newcommand{\nix}[1]{}
\begin{eqnarray}\nonumber
\iint d \Omega_{{\bm p}_0} d \Omega_{{\bm r}_0}
[\bm p_0\cdot\bm Y_{JM}^{J+1}(\Omega_{{\bm r}_0})]\cdot
[\bm p_0\cdot\bm Y_{J'M'}^{J'+1}(\Omega_{{\bm r}_0})]^*
&=&\\\frac{4\pi}{3}p_0^2\delta_{JJ'}\delta_{MM'}.&&
\end{eqnarray}
Thus, finally we get
\begin{equation}\label{final}
\frac
 1\tau=\frac{2\pi}{\hbar}\rho\left(\tfrac{E_i-E_f}{\hbar}\right)P_0
\, |\bra \psi_f |e\phi(r_0) |\psi_i \ket|^2 \,,
\end{equation}
where
\begin{multline}
|\bra \psi_f |e\phi(r_0) |\psi_i \ket|^2 \equiv\\
\frac{1}{4 \pi} \int d \Omega_{{\bm p}_0}
\int d \Omega_{{\bm r}_0}
|\bra \psi_f |e\phi(\bm r_0,{\bm p}_0)
|\psi_i \ket|^2=\\
 \frac{e^2p_0^2}{3}
\sum\limits_{JM}
\frac{\alpha_J^2}{2J+1}\frac{1}{r_0^{2J+4}}|\bra \psi_f|r^JY_{JM}|\psi_i \ket|^2\:.
\end{multline}

\section{Discussion}\label{sec:discussion}
%In this section we will estimate by the order of magnitude the intraband
%relaxation rate of the confined electron due to the coupling with local
%vibrations of the quantum dot environment considered in the previous Section.
%
%Energy of the charge $e$ in the medium with dielectric constant $\varepsilon$
%separated by $r$ from the point dipole $p$ is of the order of
%\begin{equation}\label{energy}
% e\phi\sim \frac{ep}{\varepsilon r^2}\:.
%\end{equation}
%Energy squared and averaged over the dipole coordinate is
%\begin{equation}
% \langle e^2\phi^2\rangle=\int\limits_{R}^\infty \left(\frac{ep}{\varepsilon r^2}\right)^2
% 4\pi P_0 r^2 dr \sim
% \frac{e^2p^2P_0}{\varepsilon^2 R} \:.
%\label{quadrat}
%\end{equation}
%Here  $P_0$ is the dipole concentration, $R$ is the nanocrystal radius.

We need to estimate the mean value of the dipole moment corresponding to a given local vibrational mode.
Let us first introduce characteristic values, $u$ and $M_0$, for the mean square displacement  and atomic mass 
of the atoms participating in the vibration, respectively (we neglect the fact that we deal with different sorts of atoms). This enables us to write
\[
\sum\limits_{i=1}^N  M_i \langle u_i^{(j)\,2} \rangle\omega_{\rm ph}^{(j)\,2} \sim N  M_0 u^2 \omega_{\rm ph}^2 \,,
\]
where $N$ is the number of atoms participating in the vibration, $ \omega_{\rm ph}\equiv \omega_{\rm ph}^{(j)}$.
At low temperatures
\[
N  M_0 u^2 \omega_{\rm ph}^2\sim \hbar \omega_{\rm ph} \,,
\]
and, therefore,
\begin{equation}
u\sim\sqrt{\frac{\hbar}{N M_0 \omega_{\rm ph}}}\:.\label{u}
\end{equation}
The dipole moment characterizing this vibrational mode is related to $u$
through
\begin{equation}
 p \sim eu \gamma \,,
\label{p}
\end{equation}
where the factor $\gamma$ depends on the relative orientations of
displacements for atoms participating in the local mode.\onlinecite{gamma}

%Substituting Eqs.~(\ref{p}) and~(\ref{u}) into Eq.~(\ref{quadrat}) we get
%\begin{equation}\label{res1}
 %\frac{e^2p^2P_0}{\varepsilon^2 R}   \sim
 %\frac{e^4}{R\varepsilon^2}\frac{P_0 \hbar \gamma^2}{N  M_0\omega_{\rm ph}}\:.
%\end{equation}
The other glass parameters entering Eq.~\eqref{final} are the concentration of dipoles, $P_0$, and vibrational density
 of states, $\rho$\:. The concentration of dipoles is of the order of
\begin{equation}\label{n}
 P_0 \sim a^{-3}N^{-1},
\end{equation}
where $a$ is the characteristic atomic scale.
% (in what follows we take it of the order of the Bohr radius,
 %$a\sim \hbar^2/m_0e^2$, $m_0$ is the free electron mass).
%
%Eq. \eqref{res1} can be
%rewritten as
%\begin{equation}
 %\frac{e^4}{R\,\varepsilon^2}\frac{P_0 \hbar \gamma^2}{ N M_0\omega_{\rm ph}}
  %\sim
% \frac{e^4}{R\,a\,\varepsilon^2}\frac{\hbar\omega_{\rm ph}\, \gamma^2}{ N^2M_0\omega_{\rm ph}^2\,  a^2}\:.
%\end{equation}
%The energy $M_0\omega_{\rm ph}^2 a^2$ can be estimated as follows:
%\begin{equation}
 %M_0\omega_{\rm ph}^2 a^2  \sim  10\cdot\frac{e^2}{a}\:,
%\end{equation}
%where we used the fact that $M_0\sim 10^{5}m_0$ and
%$\hbar\omega_{\rm ph}\sim  100~{\rm meV} \sim 10^{-2}e^2/a$\:. Therefore,
%\begin{equation}\label{m2}
% \frac{e^2p^2 P_0}{\varepsilon^2 R} \sim
% 10^{-1}\cdot\frac{e^2}{N^2a}\cdot\frac{a}{R}\cdot\frac{1}{\varepsilon^2}
% \cdot\hbar\omega_{\rm ph} \cdot \gamma^2
%  \sim 1~{\rm meV}\cdot N^{-2} \gamma^2 \hbar\omega_{\rm ph}\:.
%\end{equation}
%The inverse relaxation time is of the order of
%\begin{equation}
 %\frac{1}{\tau}\sim \frac{1}{\hbar}\rho\left(\omega_{\rm ph}=\frac{E_i-E_f}{\hbar}\right)
 %\langle e^2\phi^2\rangle,
%\end{equation}
The rough estimation for the density of states is $\rho\sim1/(\hbar\omega_{\rm ph})$.
% Combining
%the result with \eqref{m2} we finally obtain
%\begin{equation}
 %\frac{\hbar}{\tau}\sim 1~{\rm meV}\cdot N^{-2} \, \gamma^2, \text{ or }
% \tau\sim N^2  \gamma^{-2} \cdot 1~{\rm ps}.
% \end{equation}
%Assuming $N \sim 10$, $\gamma \sim \sqrt{N}$ we have $\tau \sim 10~{\rm ps}$. This value
%of the relaxation time is rather short.
%However, we should also have
%in mind that the we could overestimate the value of spectral
%function $\rho (\omega)$ at the corresponding frequencies.
% Indeed,
%our approximation $\rho (\omega) \omega \sim 1$ implied that all
%the local vibrations participate in the relaxation.
Numerical simulations performed for
silica glass ~\onlinecite{taraskin1,taraskin2,olig} show that the actual phase volume of the high-frequency localized vibrations is
at least by a factor of 5 less than the total phase volume. This factor decreases the density of states by about an order of magnitude.

\section{Model calculation}\label{sec:calculation}

\begin{figure}
 \includegraphics[width=0.47\textwidth]{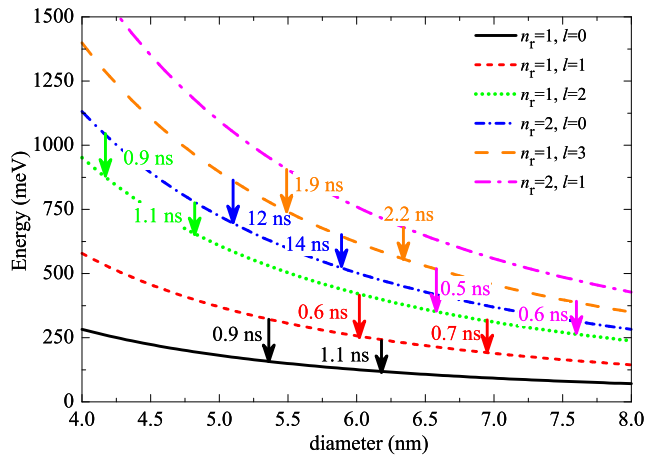}
 \caption{\label{levels}
 Relaxation time between the first six energy levels for different diameters of the nanocrystal. The values of $n_r$ and $l$ are shown for each curve. The transition time for the corresponding levels at given diameter is indicated near the vertical arrows.}
\end{figure}

In this Section we will perform a model calculation of the
relaxation time due to the coupling of electron confined in
a quantum dot with local
vibrations of the quantum dot environment. The quantum dot is treated
as a spherical quantum well with the infinite potential barrier. We consider
an electron from a simple band with an isotropic effective mass $m^*$
and use the effective mass approximation.
In this
case the electron states are characterized by the radial quantum number $n$, the orbital angular momentum $l$, its projection $m$
onto an arbitrary axis, and a projection of the electron spin.
Neglecting for simplicity the electron spin,
within the effective mass approximation
one can characterize the
electron states only by the envelope wave function
\begin{equation}\label{psi}
 \psi_{nlm}(\bm r)=\sqrt\frac{2}{R^3}Y_{lm}(\theta,\phi)\frac{j_l(\phi_{nl}r/R)}{j_{l+1}(\phi_{nl})}
\end{equation}
where $j_l(x)$ is the spherical Bessel function,
the numbers $\phi_{nl}$ are found from
\[
j_l(\phi_{nl})=0,
\]
and the energy levels
\begin{equation}\label{Enrl}
\quad E_{nl}=\frac{\hbar^2}{2m^*R^2}\phi_{nl}^2
\end{equation}
are degenerate over $m$.~\onlinecite{zeros}
Fig.~\ref{levels} shows the dependence of first six energy levels
on the nanocrystal diameter, $2R$, for $2R>4$ nm, where the density of states
mass in Si, $m^*=0.33 \, m_0$ was used ($m_0$ is the free  electron mass).
Comparison of the energy level positions with those provided by more
accurate calculations~\onlinecite{andrei1,andrei2} suggests that
the approximation of infinite
barriers is reasonable for the dot under consideration.

A transition between two of the levels of size quantization with the energies (\ref{Enrl}) is possible when their
diameter-dependent  difference falls within the
 spectral range,
  \begin{equation}\label{conserve}
120~{\rm meV} \lesssim\hbar\omega_{\rm ph}\lesssim 160~{\rm meV}\:,
\end{equation}  
  corresponding to high-frequency local vibrations of the silica glass \onlinecite{taraskin1,taraskin2,olig}.
Vertical arrows in Fig.~\ref{levels} indicate these allowed
regions and the corresponding transition time is shown near each arrow.
The calculation is carried out for the set of QD parameters close to that
of Si/SiO$_2$ system: $\varepsilon_{\rm in}=3$,
$\varepsilon_{\rm out}=12$. %\onlinecite{CardonaBook}
Our model is too simple to describe the spectral dependence of glass parameters. Thus, we take them at a fixed value in the middle of the range (\ref{conserve}): $\hbar \omega_{\rm ph}=140$ meV.  Other parameters used are as follows: $N=15$, $\gamma=\sqrt{N}$,
     $P_0=(1/N)\cdot 1.5\times 10^{22}$~cm$^{-3}$ and $\rho(\hbar\omega)=1/(5\hbar\omega_{\rm ph})$. The
      vibrating mass, $M_0$, was taken as $M_0=(1/3)M_{\rm Si}+(2/3)M_{\rm O}$ where  $M_{\rm Si}$ and $M_{\rm O}$ are
      the silicon and oxygen atomic masses, respectively. Since the levels \eqref{Enrl} are degenerate over momentum projection $m$, we have summed the relaxation rate \eqref{final} over the final states and averaged over initial ones.

    Fig.~\ref{levels} demonstrates that typical value of the relaxation time is of the order of 1 ns.
%     The rate of the transition significantly depends on the values of orbital quantum number $l$, it is faster
%     for smaller difference between the initial and final states orbital quantum numbers.  
 The transition rate
    is proportional to $1/R$, so it changes with the nanocrystal diameter slower than the energy difference
    proportional to $1/R^2$.

\section{Conclusions}\label{sec:conclusions}
We have shown that interaction of electrons confined in a non-polar quantum
dot with  local vibrations of amorphous polar environment provides an
efficient  energy relaxation channel for the ``hot'' electrons. We demonstrated
that this mechanism controls energy relaxation in Si nanocrystals in SiO$_2$
matrix with diameters in the range of $4$~--~$8$~nm, where the energy relaxation
process can proceed through emission of a single phonon.
In this case the relaxation time is in nanosecond range.  Energy relaxation
in quantum dots with smaller diameters should be controlled by multiphonon processes.
We expect  nonradiative relaxation there to be slower with
the rate comparable  to  that of  radiative intraband transitions.\cite{emrs, delerue}

This work was supported in part by the Russian Foundation for Basic Research
and by DoD under contract No.~W912HZ-06-C-0057.
A.N.P. also acknowledges the support of the ``Dynasty'' Foundation-ICFPM.

\setcounter{section}{0}\setcounter{subsection}{0}
\renewcommand{\theequation}{A\arabic{equation}}
\renewcommand{\thesection}{Appendix}

\section{}\label{A:2}
\setcounter{equation}{0}
Let us calculate electric filed induced by a point dipole $\bm p_0$ positioned at at the point $\bm r_0$ outside the sphere of radius $R$. Dielectric constants inside and outside the sphere are $\ei$ and $\eo$, respectively.
We are interested in the field inside the sphere.
Within electrostatic approximation the electric field $\bm E$ can be expressed via the scalar potential $\phi$ as
\begin{equation}
 \bm E=-\grad \phi.
\end{equation}
The Poisson equation for the potential $\phi$ reads
\begin{equation}\label{ps}
 \Delta \phi=\frac{4\pi}{\eo} \div [\bm p_0\delta(\bm r-\bm r_0)],\quad (r_0>R),
\end{equation}
and the boundary conditions at the sphere surface are
\begin{gather}\label{bc}
 \phi|_{r=R-0}=\phi|_{r=R+0},\\
  \ei\frac{\partial\phi}{\partial r}\Bigl|_{r=R-0}=\eo\frac{\partial\phi}{\partial r}\Bigl|_{r=R+0}\:.\nonumber
\end{gather}
In the case of homogeneous medium ($\ei=\eo$) the solution of \eqref{ps} is obvious,
\begin{equation}\label{phi0}
 \phi_0(\bm r)=-\frac1\eo \div\frac{\bm p_0}{|\bm r-\bm r_0|}\:.
\end{equation}
In what follows it is convenient to use the basis of vector spherical harmonics
$\bm Y_{JM}^L(\theta,\phi)$~\onlinecite{vmk}\:.
The identity
\begin{equation*}
\begin{split}
 \frac{\bm p_0}{|\bm r-\bm r_0|}=\sum\limits_{JLM}\frac{4\pi}{2L+1}\frac{r_<^L}{r_>^{L+1}}
\bm Y_{JM}^L(\bm n)[\bm Y_{JM}^{L*}(\bm n_0)\cdot \bm
p_0],
\end{split}
\end{equation*}
where $r_>=\max(r,r_0),r_<=\min(r,r_0), \bm n=\bm r/r,\bm n_0=\bm
r_0/r_0$
is useful to present \eqref{phi0} as a sum of scalar spherical
harmonics $Y_{JM}$.
After calculating the derivatives we obtain
\begin{equation}
\phi_0(\bm r)=\begin{cases}
            \sum\limits_{JM}\dfrac{r^J}{r_0^{J+2}}Y_{JM}(\bm n)\xi_{JM},\quad r<r_0\\
                \sum\limits_{JM}\dfrac{r_0^{J-1}}{r^{J+1}}Y_{JM}(\bm n)\eta_{JM},\quad r>r_0
              \end{cases}
\end{equation}
with
\begin{gather}\label{xi}
 \xi_{JM}=\frac{4\pi}{\eo}\sqrt{\frac{J+1}{2J+1}}[\bm p_0\cdot\bm Y_{JM}^{J+1}(\bm n_0)]^*,\\\quad
 \eta_{JM}=\frac{4\pi}{\eo}\sqrt{\frac{J}{2J+1}}[\bm p_0\cdot \bm Y_{JM}^{J-1}(\bm n_0)]^*\:.
\end{gather}

When $\ei\ne\eo$ the scalar potential can be written as
\begin{equation}\label{phi2}
 \phi(r)=\begin{cases}
           \sum\limits_{JM}r_{JM}Y_{JM}(\bm n)\dfrac{R^J}{r^{J+1}} +\phi_0(\bm r), &(r>R)\\
       \sum\limits_{JM}t_{JM}Y_{JM}(\bm n)\dfrac{r^J}{R^{J+1}} , &(r<R)\:.
         \end{cases}
\end{equation}
Substituting \eqref{phi2} and \eqref{xi} into the boundary conditions \eqref{bc} we find the coefficients
$ t_{JM}$ and $ t_{JM}$:
\begin{gather}
 t_{JM}=\xi_{JM}\frac{R^{J+1}}{r_0^{J+2}}\frac{(2J+1)\eo}{\ei J+\eo(J+1)},\\
r_{JM}=\xi_{JM}\frac{R^{J+1}}{r_0^{J+2}}\frac{J(\eo-\ei)}{\ei J+\eo(J+1)}\:,
\end{gather}
which allows to express the scalar potential as \eqref{phi}\:.
%\acknowledgements {This work was supported by RFBR and the ``Dynasty'' Foundation -- ICFPM.}
%\vspace{1cm}


\begin{thebibliography}{20}
\bibitem{lowT}As the characteristical energies of local vibrations $\hbar\omega_{\rm ph}^{(j)}\sim 140~$meV\cite{olig}, the high-temperature range is out of interest\:.
\bibitem{gamma} In the limiting case when the dipole characterizing a given vibrational mode can be represented as $\sim N$ elementary dipoles oscillating in phase, $\gamma\sim N$. We expect that randomization of phases would lead to  $\gamma\sim\sqrt{N}$.
\end{thebibliography}
\end{document}